\documentclass[aps,showpacs,preprint,floatfix,superscriptaddress]{revtex4-1}
\usepackage{graphicx}
\usepackage{amsmath}
\usepackage{amssymb}
\usepackage{enumitem}
\usepackage{hyperref}

\usepackage{color}

\newcommand{\e}[1]{\text{e}^{#1} }

\begin{document}

\title{\Large{Bacteria display optimal transport near surfaces}\\
\large Bacteria as intermittent active chiral particles: \\
trapped by hydrodynamics, escaping by adhesion}


\author{Emiliano Perez Ipi\~na} 
\email{equally contributed}
\affiliation{Universit{\'e} C{\^o}te d'Azur, Lab. J.~A.~Dieudonn\'{e},~UMR~7351~CNRS,~Parc Valrose,~F-06108~Nice~Cedex~02,~France}
\author{Stefan Otte} 
\email{equally contributed}
\affiliation{Universit{\'e} C{\^o}te d'Azur, Lab. J.~A.~Dieudonn\'{e},~UMR~7351~CNRS,~Parc Valrose,~F-06108~Nice~Cedex~02,~France}
\author{Rodolphe Pontier-Bres} 
\affiliation{Centre Scientifique de Monaco (CSM),  8 Quai Antoine 1er, Monaco 98000, Principality of Monaco}
\author{Dorota Czerucka} 
\affiliation{Centre Scientifique de Monaco (CSM),  8 Quai Antoine 1er, Monaco 98000, Principality of Monaco}
\author{Fernando Peruani}  
\email{Corresponding author: peruani@unice.fr [or Fernando.Peruani@univ-cotedazur.fr]}
\affiliation{Universit{\'e} C{\^o}te d'Azur, Lab. J.~A.~Dieudonn\'{e},~UMR~7351~CNRS,~Parc Valrose,~F-06108~Nice~Cedex~02,~France}

\begin{abstract}
The near-surface swimming patterns of bacteria are strongly determined by the hydrodynamic interactions between bacteria and the surface, 
which trap bacteria in smooth circular trajectories that lead to inefficient surface exploration.  
Here, we show by combining experiments and a data-driven mathematical model that surface exploration of enterohemorrhagic {\it Escherichia coli} (EHEC) -- a pathogenic strain of {\it E. coli} 
causing serious illnesses such as bloody diarrhea -- 
results from a complex interplay between motility and transient surface adhesion events.  
%
These events allow EHEC to break the smooth circular trajectories and regulate their transport properties by 
the use stop-adhesion events that lead to  a characteristic intermittent motion on surfaces. 
%
%
We find that the experimentally measured frequency of  stop-adhesion events in EHEC is located at the value predicted by the developed mathematical model that  maximizes bacterial surface diffusivity. 
We indicate that these results and the developed model apply to other bacterial strains on different surfaces, which suggests 
that swimming bacteria use transient adhesion to regulate surface motion.   
%
%
%
\end{abstract}

\maketitle

The swimming of \textit{E. coli} has been the focus of a great deal of research 
that set the basis of the canonical picture of peritrichous bacterial motion in the bulk~\cite{berg1972chemotaxis,berg1993,Berg2008}.  
Bulk exploration is performed by alternating periods 
during which bacteria ``run" in roughly straight paths and ``tumbling" events that lead to abrupt changes in the moving direction.  
In the classical picture of  bulk exploration of {\it E. coli}~\cite{berg1972chemotaxis,berg1993,Berg2008,weis1990}, 
both, the distribution of run times and  durations of tumbling events are exponentially distributed.  
%
Biased motion towards chemoattractant sources has been well reported  in {\it E. coli} and it has been found to rely on the capacity of bacteria of altering its moving direction by regulating the frequency of tumbling events, in particular by decreasing the tumbling frequency when heading in the direction of the chemoattractant gradient~\cite{Schnitzer1993,Tindall2012,Cates2012,Flores2012}. 

Surface exploration of individual bacteria, on the other hand, remains comparatively less understood. Let us stress that  individual surface 
exploration should not be confused with colony expansion, commonly called swarming, which is a collective phenomenon typically studied at air-agar interfaces~\cite{Zhang2010,Peruani2012,Ariel2015}. 
As peritrichous swimmers such as {\it E. coli} move from the bulk towards a surface, 
surface-induced hydrodynamic interactions impose a series of physical constraints: 
a) bacteria  get  attracted towards the surface, with    
bacterial motion becoming effectively two-dimensional for  long periods of time~\cite{Berke2008,dileonardo2011,Drescher05072011,spagnolie_lauga_2012,schaar2015,vigeant1997}, 
b) abrupt changes in the moving direction are strongly suppressed~\cite{Molaei2014,molaei2016}, 
and c) a hydrodynamic-induced torque (together with above-mentioned surface attraction) traps bacteria in circular, almost deterministic,  trajectories~\cite{Frymier1995,Lauga2006,lauga2009hydrodynamics,li2008,Elgeti2015,Hu2015} that lead to inefficient surface exploration. For a very recent and detailed study on near-surface swimming of {\it E. coli}, see~\cite{bianchi2017}. 

Here, we show that in experiments with highly virulent, highly adhesive enterohemorrhagic {\it E. coli} (EHEC) 
surface exploration is characterized by the intermittency between run and stop phases, with the later involving surface adhesion. 
These stop-adhesion events allow EHEC to escape from the above-mentioned 
circular trajectories.  
Note that for these bacteria, surface adhesion is key for host-tissue invasion~\cite{Clements2012} 
and biofilm formation~\cite{pratt1998,jintae2006,conrad2012,conrad2011flagella}. 
Our study reveals that surface adhesion also plays a crucial role in surface exploration. 
More specifically, we show that the temporal statistics of the stop and run phases is consistent with a three-state model, with two of those states connected to stop phases and at least one of them involving surface adhesion.  
Based on this three-state model, we build a spatiotemporal motility model. The analysis of this model reveals that the experimentally measured 
frequency at which stops occur is located at the theoretically predicted optimal frequency that maximizes bacterial diffusivity. We discuss the consequences of this remarkable finding.

\begin{figure*}[!ht]
	\includegraphics[width=1.0\textwidth]{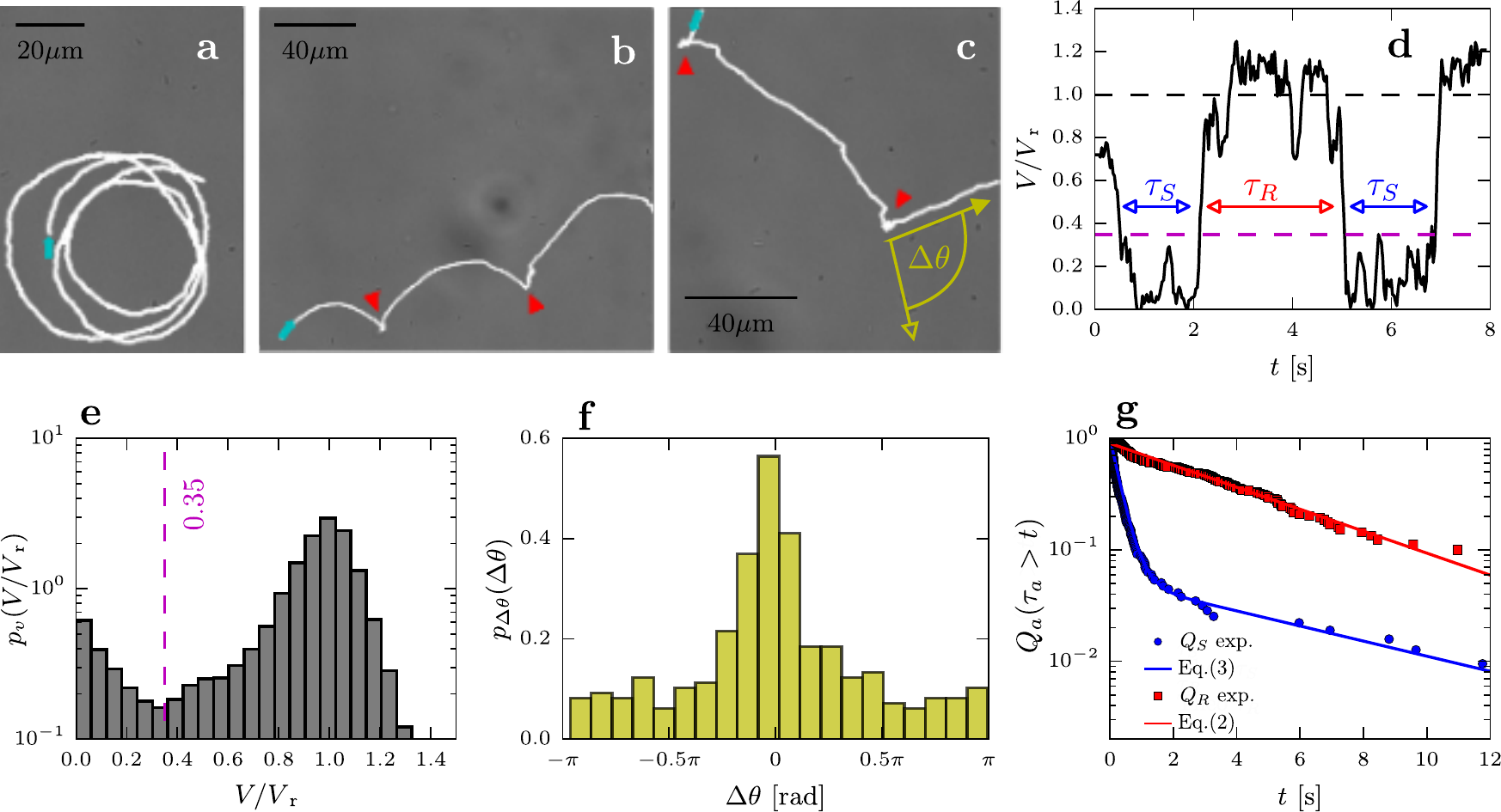}
	\caption{Experimental trajectories and their statistics. {\bf a}-{\bf c}, EHEC  display smooth circular trajectories (white curves) [{\bf a}] that are 
	interrupted by stop events (indicated by red triangles) that lead to abrupt changes in the speed ($V$) and reorientations ($\Delta \theta$) [{\bf b} and {\bf c}]. 
	{\bf d},  The characteristic intermittent behavior of the speed $V(t)$ over time $t$, whose value alternates between $V_r$ and $0$. 
	Run-time ($\tau_R$) and stop-time ($\tau_S$) periods are indicated, see text. 
	{\bf e}, Speed distribution $p_v(V/V_r)$. Its bimodal nature allows using the local minimum as a threshold,  $V/V_r=0.35$.
	{\bf f}, Distribution of changes in the moving direction (between stop and run phases), see text, $p_{\Delta\theta}(\Delta \theta)$. 
	{\bf g}, Survivals curves $Q_a(\tau_a>t)$ with $a=R$ (red squares) and $a=S$ (blue circles); the solid curves correspond to the theoretical prediction given by Eq.~(\ref{eq:CDF_Run}) and Eq.~(\ref{eq:CDF_Stop}). 
	See SI for more details and movies.}
	\label{fig:Experiments}
\end{figure*}

\section{Experiments}
We study the behavior of EHEC -- strain EDL931, serotype O157:H7, which contains a large number of adhesins, including Type 1 fimbria~\cite{EHECAdhesin} --  near the bottom surface of an {\it invitrogen Attoflour}\textsuperscript{\textregistered}  chamber 
filled with 4~mm height liquid film of DMEM medium at 37$^\circ$C. 
Using phase-contrast microscopy at 40x magnification, we record the behavior of bacteria near the glass coverslip at 30 frames per second.   
Figs.~\ref{fig:Experiments}{\bf a}-{\bf c} show examples of cell trajectories (see Movie S1, S2, and S3). 
Bacteria display smooth, typically circular trajectories that are interrupted by stop events (indicated by red triangles) that lead to abrupt changes in the speed $V$ (Fig. ~\ref{fig:Experiments}{\bf d}) and moving direction ($\theta$), see \ref{fig:Experiments}{\bf b}-{\bf c} and Movie S4. 
Note that we express  the velocity vector as $V(t)\hat{\mathbf{e}}(\theta(t))$, with $\hat{\mathbf{e}}(\cdot) = (\cos(\cdot), \sin(\cdot))$ a unit vector.
The speed distribution exhibits a bimodal shape, as shown in Fig.~\ref{fig:Experiments}{\bf e}. 
The local minimum displayed by the speed distribution is located at $V/V_{r}\approx0.35$, where $V_{r}$ corresponds
to the speed at which a local maximum is observed in the distribution (with an average $V_{r} \approx 26\mu \text{m/s}$). 
We make use of the bimodal nature of the distribution to define stop phases as those periods where $V/V_{r}\leq 0.35$ 
and run phases when $V/V_{r}>0.35$. 
This allows us to define the changes in moving direction $\Delta \theta$ as result of a stop phase as the 
difference between the value of $\theta$ immediately before and immediately after the stopping phase. 
The distribution $p_{\Delta\theta}(\Delta \theta)$, shown in Fig.~\ref{fig:Experiments}{\bf f}, is centered around $0$ 
and $p_{\Delta\theta}(\Delta \theta)>0$ in the whole interval $\Delta \theta \in [-\pi,\pi]$.
%
The distribution of times associated to  run ($\tau_{R}$) and stop ($\tau_{S}$)  phases 
is presented in Fig.~\ref{fig:Experiments}{\bf g} in the form of survival curves $Q_a(\tau_a>t)$, with $a = \{R,S \}$, i.e. the probability of observing for instance a running time $\tau$ greater than $t$. 
These quantities play a fundamental role in our analysis. 
We find that $Q_R(\tau_R>t)$ is consistent with an exponential decay with characteristic time of $\langle \tau_R \rangle \approx 4.4$s, which lets us define the stop frequency as $1/\langle \tau_R \rangle$. 
%
 %
 While the statistics of $Q_R(\tau_R>t)$ suggests a Poisson process regulating the run-time duration, a scenario qualitatively consistent with bulk motion~\cite{berg1972chemotaxis,Berg2008},    
 the survival curve $Q_S(\tau_S>t)$ indicates that the stop phase is more complex.  
In particular, we find that $Q_S(\tau_S>t)$ is of the form $A \exp[-a\,t] + B \exp[-b\,t]$ with an  average stop duration  $\langle \tau_S \rangle \approx 0.49\,\text{s}$, which is 
 about 63~\% larger than the -- strain-dependent -- average duration of a tumbling event for non-EH {\it E. coli}, which is $\approx 0.3$~s and whose distribution is a single exponential~\cite{turner2016visualizing}. 

\begin{figure}
	\includegraphics[width=1.0\columnwidth]{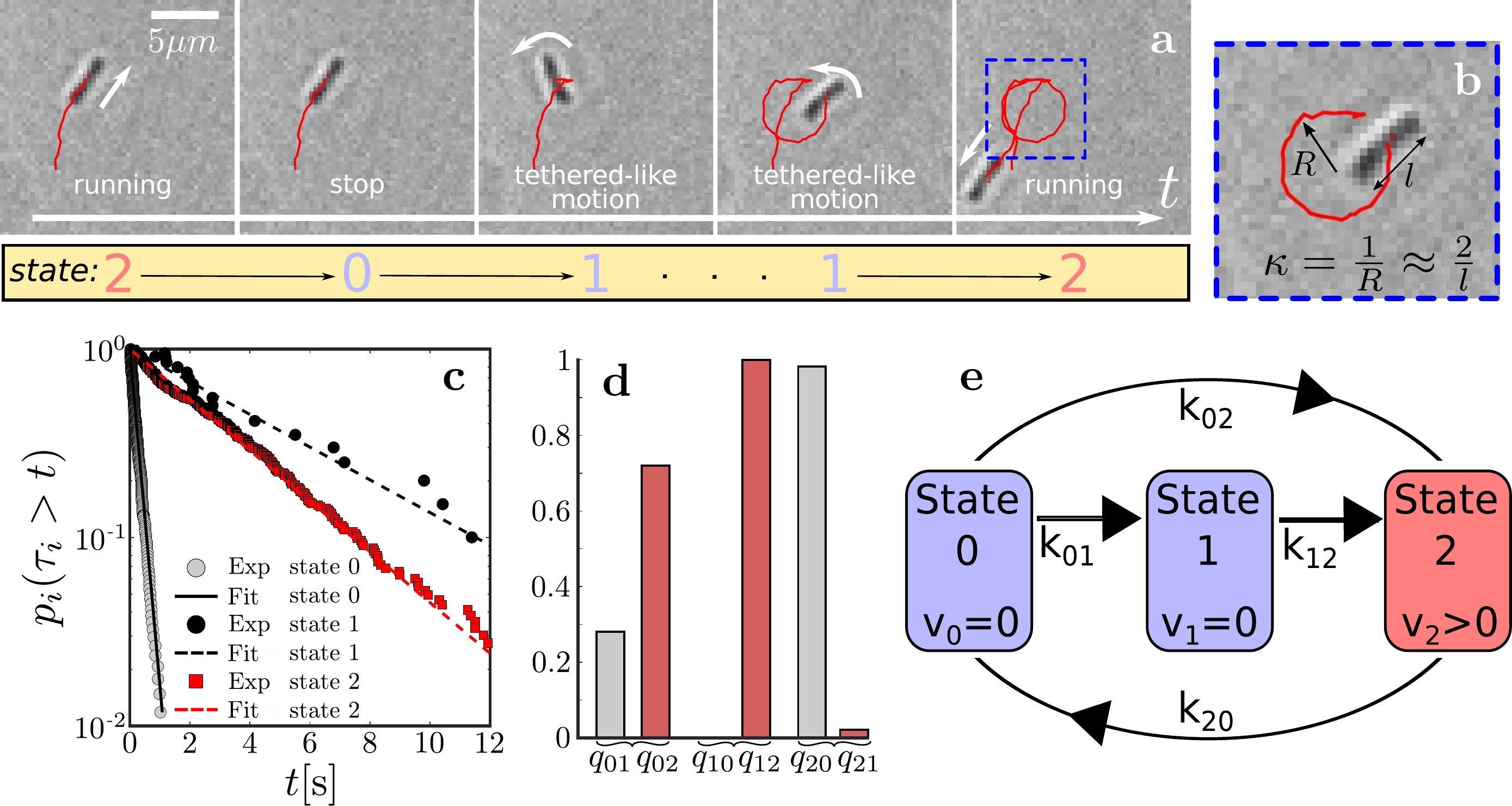}
	\caption{Evidence of three states. {\bf a}, Example of the evolution of a bacterium that undergoes a transition from a running phase (state 2) to a stop phase not  involving tethered-like motion (state 0) that transitions to  
	a tethered state (state 1) to finally switch again to a running phase. The bacterium trajectory is shown by the red curve. See Movies S5 and S6. {\bf b}, It shows that tethered-like motion can be identified by looking at the curvature $\kappa$ of the trajectory that is of the order of $1/\ell$, with $\ell$ the length of the bacterium. {\bf c}, It shows the survival curves $p_i(\tau_i>t)$ associated to state $i$. {\bf d}, The conditions probabilities $q_{i\,j}$ that the bacterium transitions from state $i$ to $j$, see text. 
	{\bf e}, The resulting Markov chain model that describes that describes the evolution of the bacterium's state. Transition rates $k_{i\,j}$ are estimated from {\bf c} and {\bf d} as explained in the text. For transition rate values, see text.}
	\label{fig:circuit}
\end{figure}

\section{Experimental evidence of three behavioral states}  
The bimodal speed distribution,  Fig.~\ref{fig:Experiments}{\bf e}, reflects an intermittent dynamics in $V(t)$,  
where the speed is either fluctuating around $V_{r}$  or around zero, Fig.~\ref{fig:Experiments}{\bf d}.  
The dichotomic dynamics displayed by $V(t)$ may lead us to think that the observed phenomenon is consistent with two behavioral states, 
one associated with $V\approx V_{r}$ and another one with $V\approx 0$. 
However, the fact that  
$p_S(\tau_S>t)$ is the sum of two exponentials suggests  the existence of two behavioral states associated with the stop phase, and thus to the presence of a minimum of three behavioral states, two associated to the stop phase and one to the run phase.    
Further inspection of the data reveals that   during periods when  $V(t)/V_{r}\leq0.35$  it is often observed that the bacterium is not at rest but rotating around one of the tips of the bacterium's body:  
a motion that is analogous to the rotations observed in experiments with tethered bacteria~\cite{1974Natur.249...73S,Berg2008}, providing 
clear evidence that these stops events  involve  adhesion  to the surface, see  Fig.~\ref{fig:circuit}{\bf a} and {\bf b}  and SI for movies. 
This type of tethered-like motion can be easily detected by monitoring   the local curvature $\kappa(t)=1/R(t)$ of the bacterial trajectory. 
When the trajectory corresponds to circles or circular arcs of a radius comparable to the length ($\ell$) of the bacterium -- in practice we use $1/3 <R/\ell<2$ -- the bacterium is in a tethered state.  
Tethered motion is typically not continuous, but intermittent: periods of rotational motion and no motion alternate each other. The detaching of the flagellar bundle from the surface is evidenced when the bacterium resumes free swimming and $V(t)/V_{r}>0.35$. 
Thus, we define the following states: ``0" indicates that the bacterium is not moving -- i.e. $V(t)/V_{r}\leq0.35$ -- and in a non-tethered state, ``1" corresponds to the bacterium in a tethered state, and ``2" implies that the bacterium is running, $V(t)/V_{r}>0.35$.  
Provided these definition, we can estimate from the experiments the probability of finding the bacterium in state 0, 1, or 2, denoted $p_i$ with $i \in \{0, 1, 2\}$. Fig.~\ref{fig:circuit}{\bf c} shows that the corresponding survival curves are exponential.    
Furthermore, we can compute from the data the conditional probabilities $q_{i\,j}$ for a particle in state $i$ to transitions to state $j$. 
Since from $i$, there are two possible transitions, e.g. $j$ and $k$, then $q_{i\,j}+q_{i\,k}=1$.  
Thus, from Fig.~\ref{fig:circuit}{\bf c} and {\bf d}  all transitions rates $k_{i\,j}$ from state $i$ to $j$  can be estimated as follows. Knowing the average time in state $i$, which is $\mathcal{T}_i=\left[ k_{i\,j}+k_{i\,k} \right]^{-1}$ (Fig.~\ref{fig:circuit}{\bf c}),  
and  $q_{i\,j}$ and $q_{i\,j}$ (Fig.~\ref{fig:circuit}{\bf d}), the rates are computed as $k_{i\,j} = q_{i\,j}\,\mathcal{T}_i^{-1}$~[[Note that in original publication, Perez-Ipi{\~n}a et al. Nature Physics {\bf 15} 610-615 (2019), this expression was reported with a typo: the ``-1" exponent was missing]]. 
We find that  97\% of transitions from  2 occur to state 0, and thus for simplicity in the following we neglect transitions $2 \to 1$.  
In addition, since the only detectable transition out of state 1 -- i.e. involving surface detachment is by observing that the bacterium resumes swimming, i.e. transitions to 2  --  the transition $1 \to 0$ is not present. 
Following the above described procedure, we obtain: $k_{01} = (0.20 \pm 0.02)$~s\textsuperscript{-1}, $k_{02} = (3.71 \pm 0.2)$~s\textsuperscript{-1}, $k_{12} = (0.21 \pm 0.02)$~s\textsuperscript{-1}, and $k_{20}= 0.22 \pm 0.05$~s\textsuperscript{-1}. 
Note that it is possible to conceive alternative statistical treatments that involve 6 transition rates that despite the added complexity lead to the same results, see SI.  
In summary, we find that the observed behavior can be accurately described by the simple  Markov chain shown in Fig.~\ref{fig:circuit}{\bf e}, whose master equation reads:
\begin{subequations}
\label{eq:master}
\begin{align}
\label{eq:master1}
\partial_t p_0(t) &= - (k_{02}+ k_{01}) p_0 + k_{20} p_2, \\
\label{eq:master2} 
\partial_t p_1(t) &= - k_{12} p_1 + k_{01} p_0, \\ 
\label{eq:master3}
\partial_t p_2(t) &= - k_{20} p_2 + k_{02} p_0  + k_{12} p_1\,.  
\end{align}
\end{subequations}
%
%
Computing the survival curves $Q_R(\tau_R > t)$ and $Q_S(\tau_S > t)$ from these equations imply to solve first passage time probabilities on the Markov chain shown in Fig.~\ref{fig:circuit}{\bf e} 
using suitable boundary and initial conditions; for details see SI. 
We find that  
\begin{equation}
Q_R(\tau_R > t) = \e{-k_{20}t} \label{eq:CDF_Run}\,, 
\end{equation}
while $Q_S$ we obtain: 
\begin{equation}
Q_S(\tau_S > t) = \left( 1 - \frac{k_{01}}{\alpha} \right) \e{-(k_{01}+k_{02})t} + \frac{k_{01}}{\alpha} \e{-k_{12} t} \label{eq:CDF_Stop}
\end{equation}
where $\alpha = k_{01}+k_{02} - k_{12}$.  

\section{Spatial motion} 
The proposed three-state model can be used as the basis for a spatial model, which should be able to describe the large 
variety of trajectories observed in experiments and incorporate the following considerations:    

(i) Figs.~\ref{fig:Experiments}{\bf a} and {\bf b} show that a moving bacterium tends to describe circular trajectories interrupted by  stop events. 
This implies that state $2$ should describe an active chiral swimmer~\cite{Teeffelen2008,Ebbens2010,Kuemmel2013}  characterized by a speed $V_r$ and a radius of curvature $V_r/\Omega_0$. 
The value of $\Omega_0$, associated to the curvature in the running state, as well as the strength of moving direction fluctuations, denoted by $D_{\theta}$, can be directly extracted from 
individual trajectories by studying the velocity autocorrelation. Further details on this procedure are provided in SI. 
Thus, we 
 assigned to states $0$, $1$, $2$ the linear speeds $V[0]=V[1]=0$ and $V[2]=V_{r}>0$, and angular speeds $\Omega[0]=\Omega[1]=0$ and $\Omega[2]=\Omega_0$, respectively. 
Transitions between either stop state, i.e. $0$ or $1$, to state $2$ involve changes in the moving direction ($\Delta \theta$) as indicated in Figs.~\ref{fig:Experiments}{\bf b} and {\bf c}. 

(ii) We encode the details on the associated changes in the moving direction due to transitions $0 \to 2$ and $1 \to 2$ 
into the probability densities $g_{02}(\theta, \theta')$ and $g_{12}(\theta, \theta')$, respectively, 
where $\theta'$ denotes the moving direction before the stop event and $\theta$ after it, with $\Delta\theta=\theta - \theta'$. 

\begin{figure*}
	\includegraphics[width=1.0\textwidth]{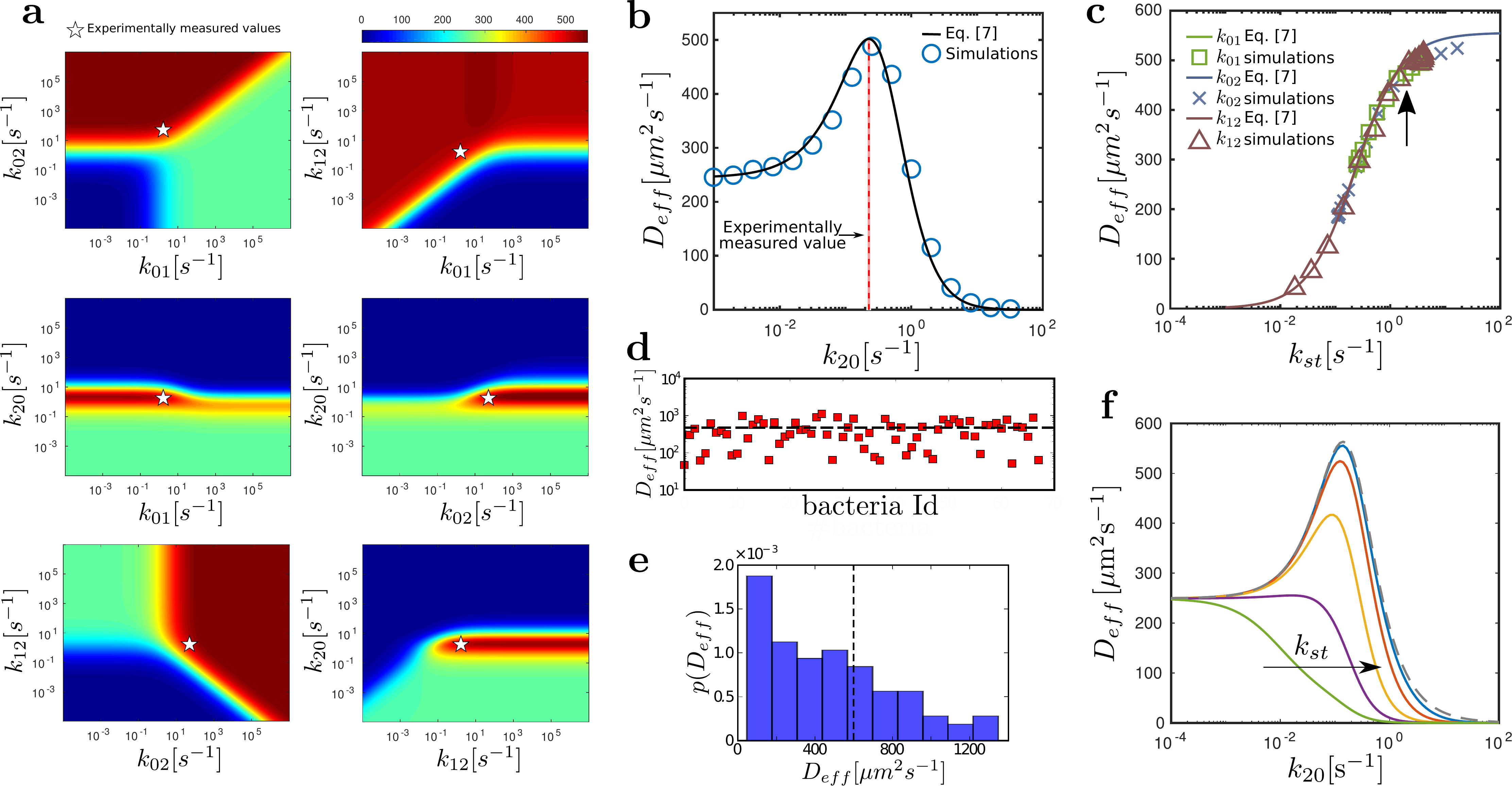}
	\caption{Diffusion coefficient. 
	{\bf a}, The value of $D_\text{eff}$, color coded and measured in units of $\mu m^2\text{s}^{-1}$, as function of two transition rates, $k_{ij}$ and $k_{i'j'}$. 
	The star in each panel indicates where the experimental value is located. 
	{\bf b}, $D_\text{eff}$ vs the stop frequency $k_{20}$. Symbols corresponds to agent-based simulations, the solid black curve to Eq.~(\ref{eq:Deff_exact_3ST_main}), and the vertical 
	red line indicates the experimentally measured value of $k_{20}$. 
	{\bf c}, $D_\text{eff}$ vs the inverse of the average stopping time given by $k_{st}=\frac{k_{01}+k_{02}}{1 + k_{01}/k_{12}}$. Symbols corresponds to agent based simulations obtained 
	by varying   $k_{12}$ (triangles), $k_{01}$ (squares), and $k_{02}$ (crosses), while the solid curve corresponds again to Eq.~(\ref{eq:Deff_exact_3ST_main}). The vertical arrow indicates experimentally obtained $k_{st}$ value. 
	Note that $D_\text{eff}$ does not exhibit a local maximum of any value of $k_{st}$.	
	{\bf d}, Experimentally estimated values of $D_\text{eff}$, where the horizontal dashed lines corresponds to the average. {\bf e}, It shows the 
	histogram of the $D_\text{eff}$ values displayed in {\bf d}, where the vertical red dashed  line corresponds to the average $D_\text{eff}$.  
	{\bf f}, $D_\text{eff}$ vs $k_{20}$ for $k_{st}$ values $0.010$,  $0.056$, $0.316$, $1.778$ and $10^4\,\text{s}^{-1}$ in the direction indicated by the horizontal arrow.
	Note that for low $k_{st}$ values, there is no optimal $k_{20}$. 	
	}
	\label{fig:Modeling}
\end{figure*}

Under these considerations, 
the spatiotemporal evolution of the three-state bacterium can be expressed via 
the probability density $p_i(\vec{x},\theta,t)$  
of finding the bacterium in state $i$ (with $i \in \{0,1,2\}$) at position $\vec{x}$ and moving in direction $\theta$ at time $t$ 
by the following continuum-time forward Kolmogorov equation with transition-jumps:  
%
\begin{subequations}
\label{eq:sm}
\begin{align}
\partial_t p_0(\vec{x},&\theta,t) = - (k_{02}+ k_{01}) p_0 + k_{20} p_2, \label{eq_m:FPE_st0}\\ 
\partial_t p_1(\vec{x},&\theta,t) = - k_{12} p_1 + k_{01} p_0, \label{eq_m:FPE_st1}\\ 
\nonumber
\partial_t p_2(\vec{x},&\theta,t) = -V_r\! \nabla(\hat{\mathbf{e}}(\theta)p_2)\! -\! \Omega_0 \partial_{\theta} p_2\! +\! D_{\theta}\partial_{\theta\theta} p_2\!-\!k_{20} p_2 \\
 &+\! k_{02}\!\! \int\! g_{02}(\theta, \theta')p_0 d\theta'\! +\! k_{12}\!\int\! g_{12}(\theta, \theta')p_1 d\theta'\!. \label{eq_m:FPE_st2}
\end{align}
\end{subequations}
Note that the spatiotemporal model defined by Eqs.~[\ref{eq_m:FPE_st0}-\ref{eq_m:FPE_st2}] 
can also be expressed by the following set of stochastic differential equations:
\begin{equation}
	\label{eq:Langevin}
		\begin{array}{cc} 
		\dot{\vec{x}}  =\! V[i(t)] \hat{e} (\theta(t))\,;  &\!\! \dot \theta  = \Omega[i(t)]\! +\! \sqrt{2 D[i(t)]}\, \xi_\theta(t)\! +\! \zeta(t) \, ,
		\end{array}
\end{equation}
where $i(t)$ denotes the state of the bacterium at time $t$, 
$\Omega[i]$ is defined as $\Omega[0]=\Omega[1]=0$ and  $\Omega[2]=\Omega_0$, $D[i]$ as 
$D[0]=D[1]=0$ and $D[2]=D_{\theta}$, $\xi_\theta(t)$ is a white noise of unit variance, and 
$\zeta(t)$ is a shot noise $\zeta(t) = \sum_{k} \beta_k \delta(t - t_k)$, 
with $t_k$ denoting the time at which the $k$-th transition from either $0 \to 2$ or $1 \to 2$ occurred and $\beta_k$ the associated  
change in  $\theta$ at that given transition. The temporal evolution of $i(t)$ as well as the times $t_k$ are dictated by 
the temporal three-state model defined above (Eq.~(\ref{eq:master})).

In order to solve Eq.~(\ref{eq:sm}), we need to specify $g_{02}$ and $g_{12}$. 
From Fig.~\ref{fig:Experiments}{\bf f}, it becomes evident that $g_{02}(\theta, \theta') = g_{02}(|\Delta\theta|)$ and the same argument applies to $g_{12}$. 
To simplify the calculations, we assume that   $g_{02}$ and $g_{12}$ are uniformly distributed in the intervals $-\phi_0\leq \Delta \theta \leq \phi_0$ and 
$-\phi_1\leq \Delta \theta \leq \phi_1$, respectively, and zero otherwise. Note that Eq.~(\ref{eq:sm}) can be solved without this approximation via a longer calculation that leads almost exactly the same result, see SI.
We fix the values of $\phi_1=\pi$ and use  $\phi_0$  to match the variance of the experimental distribution $p_{\Delta\theta}(\Delta \theta)$, Fig.~\ref{fig:Experiments}{\bf f}, 
taking into account that a fraction $q_{02}=\frac{k_{02}}{k_{01} + k_{02}}$ of the transitions from stop to run phase occur through  the path $0 \to 2$, 
while the rest ($1-q_{02}$) via $0 \to 1 \to 2$. We obtain $\phi_0 = 0.72\pi$ and $\phi_1 = \pi$.  
With the above expression for $g_{02}$ and $g_{12}$, we 
can analytically obtain 
 an effective diffusion coefficient $D_{eff}$ for 
the proposed three-state bacterial model. 
We use the Taylor-Kubo formula to express $D_{eff} =  \lim_{t\to \infty} \frac{1}{4t} \int_0^t dt' \int_0^t dt'' \langle \mathbf{v}(t') \mathbf{v}(t'') \rangle$, 
which reduces the complexity of the problem to compute $C(t',t'')=\langle \mathbf{v}(t') \mathbf{v}(t'') \rangle = \langle V[i(t')] \hat{\mathbf{e}}(\theta(t') V[i(t'')] \hat{\mathbf{e}}(\theta(t'') \rangle$, which 
can be expressed as:
\begin{eqnarray}
\label{eq:jointProb}
&&C(t',t'') =\\
\nonumber
&& \!\int \! d\theta'\!\! \int \! d\theta'' \! \sum_{i', i''} \! V[i'] V[i'']\! \cos(\theta'-\theta'') p(i'\!,\!\theta',\!t',\!i'',\!\theta'',\!t'') \,
\end{eqnarray}
where $i'$ and $i''$ refer to the state of the bacterium at time $t'$ and $t''$. 
Note that $p(i'\!,\!\theta',\!t',\!i'',\!\theta'',\!t'')=p(i'\!,\!\theta',\!t'|\!\,i'',\!\theta'',\!t'')p_{i''}(\theta'',\!t'')$.  
Both, $p(i'\!,\!\theta',\!t'|\!\,i'',\!\theta'',\!t'')$ and $p_{i''}(\theta'',\!t'')$ can be directly obtained from Eq.~(\ref{eq:sm}) 
after integrating over space Eq.~(\ref{eq:sm}).  
The computation of  $p(i'\!,\!\theta',\!t'|\!\,i'',\!\theta'',\!t'')$ requires solving Eq.~(\ref{eq:sm})  for $t\,\geq\,t''$ with initial condition $p_k(\theta, t=t'')=\delta(\theta-\theta')$ for $k=i''$ and $p_k(\theta, t=t'')=0$ for $k\neq i''$. 
The details on the calculation are provided in the SI. 
Following the described procedure, we find that: 
\begin{align}
\label{eq:Deff_exact_3ST_main}
D_\text{eff} =& V_r^2 \frac{\left( D_{\theta} + k_{20} \hat{\phi} \right)}{ 2\left(\left( D_{\theta} + k_{20} \hat{\phi} \right)^2 + \Omega_0^2\right)} \frac{k_{st}}{k_{st}+k_{20}}, 
\end{align}
where $\hat{\phi}=\left( q_0\phi_0^2 + q_1\phi_1^2 \right)/6$, $k_{st}$ is defined as $k_{st} = \frac{k_{01}+k_{02}}{1 + k_{01}/k_{12}}$, and $V_r = 26~\mu$m\,s\textsuperscript{-1}, $\Omega_0 = 0.3$~s\textsuperscript{-1} and $ D_{\theta} = 0.07$~s\textsuperscript{-1} are extracted 
from the experimental trajectories as explained in Materials and Methods and SI. 
The other parameters  are obtained from the experimental data by the procedures explained above; values have been also provided (see previous sections).

\section{Dependency  on the transition rates} 
Fig.~\ref{fig:Modeling}{\bf a} shows the value of $D_\text{eff}$ color coded as function of two rates $k_{ij}$, with all other parameters  kept constant to the experimentally measured values. 
In each panel, a star indicates the location of the experimentally obtained values of the rates. 
The exploration of the model let us identify the role and relevance of each rate. 
The most remarkable finding is that for large enough values of $k_{02}$ and $k_{12}$, and for the whole range of $k_{01}$,  
by varying $k_{20}$ we unveil the existence of a local maximum for $D_\text{eff}$ (also global maximum).  
From Fig.~\ref{fig:Modeling}{\bf b}, where  
symbols correspond to agent-based simulations (see SI for details) and the solid curve to Eq.~(\ref{eq:Deff_exact_3ST_main}), 
it is evident that $\left. \frac{dD_\text{eff}}{d k_{20}}\right|_{k^*_{20}}\!\!\!\!=0$ at an optimal $k^*_{20} \approx 0.2 s\textsuperscript{-1}$  
that is remarkably close to the value of $k_{20}$ obtained in experiments (see vertical line).   
At this value of $k_{20}$, Eq.~(\ref{eq:Deff_exact_3ST_main}) predicts a diffusion coefficient of the order of $500\,\mu\text{m}^2\text{s}^{-1}$, which is close to the average of experimentally obtained values, see Fig.~\ref{fig:Modeling}{\bf d} and {\bf e}. 
%
On the other hand, Fig.~\ref{fig:Modeling}{\bf c} shows that $D_\text{eff}$ does not exhibit a local maximum by varying $k_{st}$, which is the inverse of 
the average stopping time that is function of all other rates: $k_{01}$, $k_{02}$, and $k_{12}$.
%
Interestingly, the value of $k_{st}$ controls the potential existence of a local maximum with respect to $k_{20}$ as shown in Fig.~\ref{fig:Modeling}{\bf f}. 
This means that depending on the values of  $k_{01}$, $k_{02}$, and $k_{12}$, that may depend on the properties of  bacterial adhesins on the specific surface, 
$D_\text{eff}$ may not exhibit a local maximum and the global maximum may be trivially located at $k_{20} \to 0$, indicating that stop-adhesion events, in this case, may not help to enhance $D_\text{eff}$. 
%

\section{Concluding remarks} 
We have shown that in EHEC, individual surface exploration results 
from a complex interplay between motility and stop events leading to surface adhesion. 
These stop-adhesion events allow EHEC to break the hydrodynamic-induced circular trajectories and explore the surface by performing alternating run and stop-adhesion phases. 
More specifically, we showed that the experimental data is consistent with an arguably generic three-state  model that suggests that 
surface exploration by swimming peritrichous bacteria can be strongly enhanced by  performing transient surface adhesion events.   
In particular, the analysis unveiled that the average frequency at which stops occur in experiments (referred to as stop frequency) is located at the optimal value that maximizes bacterial diffusivity predicted by the developed theory.  
While this finding may appear as a coincidence, we found that in the experiments reported by Sauer {\it et al.}~\cite{Sauer2016} with uropathogenic {\it E. coli}  -- which importantly possess different types of adhesins from the ones found in EHEC -- and where bacteria move on and adhere to mannosylated surfaces -- i.e different surfaces from the one here analyzed -- the  surface exploration statistics is perfectly consistent with the proposed three-state model. Moreover, the stop frequency measured in those experiments also coincides with the value that according to the developed model maximizes the diffusion coefficient of the bacteria; for details and figures on the analysis of the data reported in~\cite{Sauer2016}, see SI.  
This observation strongly suggests that the proposed mechanism  is not specific to EHEC, but rather a generic mechanism of bacterial surface exploration.   
Moreover, it has been reported that {\it E. coli} and other bacteria possess mechano-sensitive channels that allow them to ``know" when they are on a surface and to perform surface-sensing~\cite{Mei2012}, 
as well as to regulate the activity of the bacterial flagellar motor~\cite{Nord2017}. 
All this suggests that bacteria may be indeed able to adapt their behavior depending on the surface properties and tune the stop frequency to the optimal value that maximizes surface diffusivity. 
Note that in the absence of  information cues,  as occurs in the studied experiments, 
a higher surface diffusivity confers an advantage to bacteria to find randomly distributed colonization sites or food patches located on the surface~\cite{Adam1968}.  
In the presence of information cues, e.g. in the form of chemical gradients, bacteria have to cope with conflicting requirements of searching and localization as explained by  Clark and Grant  in~\cite{Clark2005} and in the context of search without gradients in~\cite{vergassola2007}. 
Without entering into these important issues~\cite{Clark2005,vergassola2007}, the analysis of the model suggests that  by regulating the frequency of stop-adhesion events bacteria  
can perform biased motion on surfaces (see SI). 
Undoubtedly, surface motility experiments in well-controlled chemoattractant gradients are required to clarify these intriguing issues. 
Other promising experimental directions include the use of c-di-GMP to alter the run time distribution~\cite{Gomelsky2010}, alter the adhesion properties by using a 
different substrate material~\cite{Cookson2002}  or the use of mannose as in~\cite{Nilsson2006}  to alter the average stop time.

The collected evidence, together with the recent discoveries on  bacterial surface sensing capacity, and the  observation that 
the experimental values of the tumbling frequency in three-dimensional swimming 
do not maximize bacterial 3D  diffusivity~\cite{Berg2008}, let us speculate that peritrichous bacteria have evolved 
to become optimal surface explorers by the identified exploration  
mechanism involving  transient adhesion events. This should 
not come as a surprise if we consider that nutrients in aqueous environments tend to accumulate at surfaces~\cite{Fletcher1994}.

\section{Material and Methods} 

\subsection*{Experimental setup}
\emph{Microorganisms}: Enterohemorrhagic {\it Escherichia coli} EDL931 (CIP 103571 ``Institut Pasteur'') serotype O157:H7 (EC) was kindly provided by St\'ephane M\'eresse, Facult\'e des Sciences de Luminy, Centre d'Immunologie de Marseille-Luminy (CIML), INSERM-CNRS, Marseille, France. Bacteria were stored in Luria-Bertani (LB) medium plus 15\% glycerol at $-80^{\circ}$C.

\emph{Preparation for video-microscopy:} Bacteria were grown overnight into LB broth medium without shaking (condition that preserved flagellum). Bacteria were pelleted by gentle centrifugation (2500 rpm for 10 minutes) and resuspended in DMEM medium. Subsequently, 2 ml of liquid was disposed into \emph{invitrogen Attoflou\textsuperscript{\textregistered}} cell chambers, leading to a density of approx. 10\textsuperscript{6} bacteria/dish. The circular cell chambers have a diameter of 25~mm, which lead to a heigth of approx. 4~mm of the liquid above the bottom glass surface of the cell chamber. For time-lapse video microscopy, the chambers were placed in a humidity (95\%), CO2 (5\%) and temperature (37\textsuperscript{o}C) - controlled environment. The focus was set on the cover slip at the bottom of the cell chamber, in order to record bacterial motion close to the surface glass/liquid. At least 30 minutes were given to the system to equilibrate prior to recording bacterial motion.

\emph{Records of video-microscopy:} Motile bacteria were recorded by phase-contrast microscopy using a Leica DMI6000 B inverted microscope equipped with a high-sensitive Ropper CoolSnap HQ2 CCD camera (Photometrics) at 40x magnification (numerical aperture: 0.75, Leica HCX PL Fluotar PH2). Images were acquired with the LAS-AF software (Leica, Germany) at a rate of 35 fps.  

\subsection*{Data analysis and fitting of parameters}
Videos were analyzed using \emph{fiji} platform \cite{schindelin2012fiji}. 
Trajectories of 147 individual bacteria were tracked using two different tools: MTrackJ~\cite{meijering2012methods} for manual tracking and TrackMate~\cite{jaqaman2008robust} for semi-automatic tracking. 
From the tracking, the bacteria's center of mass $\{ \vec{x}_j\}$ with $\vec{x}_j = (x(t_j), y (t_j))$ at times $t_j = j \cdot \Delta t$ for $j=1,...,N$ time steps were obtained for each bacterium, where is $\Delta t = f_\text{M}^{-1}$ the inverse of the microscope frame rate. 
The speed at time $t_j$ was computed as
\begin{equation*}
	V_{j,n} = \frac{1}{n\Delta t}  || \vec{x}_{j+\frac{n}{2}} - \vec{x}_{j-\frac{n}{2}}  ||
\end{equation*}
i.e. as the average speed over the  neighboring $n=6$ time steps in order to remove rapid speed fluctuations. 
For each bacterium, a reference speed $V_\text{r}$ was computed as the arithmetic average of a subset of speed values $\{V_s\}$ that satisfy $V_s \geq 0.5 V_{max}$, where $V_{max} = \text{max}\{ V_j \}$ is the maximum speed of the trajectory.
As was mentioned in the main text, $V_r$ was used to identify the {\it stops} events. Plotting the distribution of $V_{j,n}/V_r$ for all the trajectories we observed two local maximums, one at $V_r$ and the other at $0$. A minimum value is located at $V_{j,n}/V_r = 0.35$ and was used as a threshold to identify run and stop phases. A stop phase is defined as the consecutive times $s_i=\{t_j\}$ where the condition $V_{j,n}/V_r < 0.35$ is satisfied. In the same way, a run phase corresponds to consecutive times $r_l=\{t_k\}$ where $V_{k,n}/V_r > 0.35$. 

The moving direction $\theta (t_j)$ was computed by first smoothing the $x$ and $y$ components of the trajectory independently using a Gaussian filter with width $\sigma=3$ (in order to remove fluctuations such as wobbling motion of the bacterial body) and subsequently using the quadrant-sensitive arctangent function on the smoothed trajectory $\{ \vec{x}_j^\text{S}\}$:
\begin{equation*}
	\theta_{j} = \arctan \left( \frac{y^\text{S}_{j+1}-y^\text{S}_{j}}  {x^\text{S}_{j+1}-x^\text{S}_{j} }  \right).
\end{equation*}
The change in the moving direction during a {\it stop} event was defined as $\Delta\theta= \theta(\text{min}(r_{l})) - \theta(\text{min}(s_i))$, again where the $r_l$-event is the running event that immediately follows the $s_i$-event.

In order to estimate the population mean values of the parameters we only consider the running phases longer than $100\Delta t$. 
Then we estimated the mean value of the running speed, $\langle V_r\rangle$ as the average over all the $V_r$. In the case that a trajectory did not have any stops we re-defined $V_r$ as the average in time of $V_{j,n}$.
The angular speed, $\Omega_0$, was computed using the curvature $\kappa$ of the running phases. 
The curvature $\kappa$ can be geometrically calculated by, 
\begin{equation*}
 \kappa(t_i) = -\frac{|\ddot{\vec{x}}(t_i) \times \dot{\vec{x}}(t_i)|}{|\dot{\vec{x}}(t_i)|^3} = \frac{\dot{x}(t_i)\ddot{y}(t_i) - \dot{y}(t_i)\ddot{x}(t_i)}{(\dot{x}^2(t_i) + \dot{y}^2(t_i))^{3/2}}
\end{equation*}
The temporal first and second derivatives appearing in this equation are estimated using finite differences.
Finally, $\Omega_0$ and $\bar \kappa$ are geometrically related by
\begin{equation}
\Omega_0 = \bar \kappa V_r,
\end{equation}
such that $\Omega_0$ can be obtained if previously $\bar \kappa$ and the mean speed $V_r$ have been measured independently. 

The angular diffusion coefficient, $D_\theta$, was computed from the (average) correlation of the moving direction, $\langle \hat{\vec{e}}_i.\hat{\vec{e}}_{i+j}  \rangle_i$, where $\hat{\vec{e}}_i= \frac{\dot{\vec{x}}(t_i)}{V_{i,n}}$ is the (average) direction of motion of the particle at time $t_i$. 
For a chiral active particle the correlation of the moving direction can be expressed as:
\begin{equation}
\label{eq:directionCorr}
\langle \hat{\vec{e}}_i.\hat{\vec{e}}_{i+j}  \rangle_i = e^{- D_\theta j \Delta t} \cos \left( \Omega_0 j \Delta t \right).
\end{equation}
Finally, knowing $\Omega_0$, it is possible to extract the values of $D_{\theta}$ fitting the data with Eq.~(\ref{eq:directionCorr}).

All fitting was done using the Levenberg-Marquardt nonlinear least-squares algorithm.  
In SI  the histograms of the estimated values of the model parameters obtained for the running events analyzed are shown.

\subsection*{Notes on the time distributions estimations}

For the analysis of the duration of the runs and stops we have to consider that we can track bacteria as long as they are in the camera frame, meaning that we cannot follow them for long times and we only see a part of their trajectories. 
This implies that we have a bias in the accessible data towards the short times. 
In the analysis of the duration of run and stop phases, in order to avoid a bias towards short times, {\it ``partial''} events were included, {\it i.e.} those events that were not observed from the beginning till the end, {\it e.g.} the partial running event of a bacterium that enters/exits the camera frame (in such cases we do not know when the running event started/finished).
These partial events were analyzed following the Kaplan-Meier method to estimate the survival distribution. 
For the running time distribution we applied the same procedure as the one used in~\cite{berg1972chemotaxis} in order to account for the individual variability.  
%
%
%
%

%
\section*{Acknowledgments}
We thank L. G{\'o}mez Nava, R. Gro{\ss}mann, A. Be'er, and Giovanni and Giorgio Volpe  for insightful comments on the text. Experiments were performed at  C3M Imaging Core Facility (Microscopy and Imaging platform C{\^o}te d'Azur, MICA). We acknowledge support from Grant ANR-15-CE30-0002-01 and from Biocodex.

\section*{Author contributions}
D.C. and F.P. designed the study. D.C. and R.P.B. performed experiments. E.P.I., S.O., and F.P. performed the image and statistical analysis of the data and derived the mathematical models used to interpret the data. F.P. wrote the manuscript with the help of all authors.

\bibliographystyle{apsrev}



\end{document}